# Fine Structure Constant in the Slope of Regge Trajectories

David Akers*
*Lockheed Martin Corporation, Dept. 6F2P, Bldg. 660, Mail Zone 6620,*
*1011 Lockheed Way, Palmdale, CA 93599*
*Email address: David.Akers@lmco.com

Recent research has indicated that meson and baryon Regge trajectories are nonlinear and that all current models are ruled out by data. Tang and Norbury have identified a number of properties for Regge trajectories: a test zone for linearity, divergence, parallelism and intersecting lines. Likewise, Inopin has reconstructed Regge trajectories for mesons and baryons, indicating that a majority of trajectories is grossly nonlinear. In most of the models for Regge trajectories, there is no indication that researchers have studied particle binding energies to explain the deviation of experimental data from theory. In terms of quark models, Mac Gregor has studied binding energies in the constituent quark (CQ) model, and we shall turn to this model for help in explaining the slope of meson Regge trajectories.
PACS number(s): 12.40.Nn, 12.39.-x

## I. INTRODUCTION

Regge trajectories have been studied for over 40 years. Tullio Regge initially introduced the concept [1, 2]. A great number of quark models have been introduced to explain the properties of Regge trajectories. According to Inopin [3], the introduction of so many models has lead to confusion, because these models are contradictory to each other. For a comprehensive review of current research in Regge theory, see the work of Inopin [3] for the many models proposed to explain the experimental data. In this paper, we propose to review an earlier model by Barut [4, 5], establish slopes for meson and baryon Regge trajectories from the data published by the Particle Data Group [6], and show how the Regge trajectories are dependent upon a 70 MeV quantum proposed by Mac Gregor [7]. In the next section, we compare our results with the recent review by Tang and Norbury [8] on the properties of Regge trajectories.

## II. MESON REGGE TRAJECTORIES

In the 1970s, many quark models were proposed before the establishment of the Standard Model of particle physics. Mac Gregor noted regularities in the hadron spectrum with 70, 140, and 210 MeV energy separations [9, 10]. He attributed these energy separations to both excitation and rotational spectra [11], and developed the idea of constituent-quark (CQ) binding energies [7]. At the same time, Barut attempted to establish a model of hadrons based upon the existence of magnetic charges and was the first to coin the term "dyonium" for the binding of two spinless dyons [4, 5]. The idea of the dyon was first proposed by Schwinger [12]. In his discussion on Regge trajectories, Barut mentioned that the proportionality of the slope of the mass formula was related to the fine structure constant or to the number 137. Barut also noted that Nambu was the first to note the proportionality of the masses to 137 [13]. Later, Nambu's empirical mass formula was derived from a modified QCD Lagrangian with Yang-Mills fields and a multiplet of scalar Higgs fields [14].

From the dependence of the mass formula on the magnetic form factor $G_M$, Barut noted that "we can understand why mesons and baryons have essentially similar Regge trajectories" [4]. In this paper, we shall establish the fact that mesons and baryons have *different* slopes for Regge trajectories, and yet these slopes are each proportional to the fine structure constant or to the number 137.

In studying the relativistic Balmer formula for the case when $\alpha$ is large, Barut obtained the following mass formula [5]:

$$M^2 = (m_1)^2 + (m_2)^2 + 2m_1m_2J/\alpha, \quad (1)$$

where α is the fine structure constant and J is the angular momentum number. In order to apply this mass formula to the case of mesons, we note that the pion would be the first low-mass meson on a Regge trajectory, such that the mass formula becomes:

$$M^2 = (m_1 + m_2)^2 + 2m_1m_2J/\alpha. \quad (2)$$

In the CQ model, Mac Gregor has noted that the pion is a composite of two 70 MeV/$c^2$ mass quanta or $m_1 = m_2 = 70$ MeV = 0.070 GeV. (For simplicity, we shall note this $m_1$ as the 70 MeV quantum, dropping the notation for the speed of light.) Thus, the slope of the meson Regge trajectories is:

$$\begin{aligned} \text{Slope (mesons)} &= 2m_1m_2/\alpha, \\ &= 2(0.070)^2(137), \quad (3) \\ &= 1.3426\ (\text{GeV}^2). \end{aligned}$$

The slope of Eq.(3) is in good agreement with the slope (1.2 $\text{GeV}^2$) derived by Andreev and Sergeenko in their paper on the relativistic quantum mechanics [15].

We now reproduce the spectra of mesons for masses below 2700 MeV which are dependent upon the slope of Eq.(3). These Regge trajectories are shown in Figures 1 to 7. The meson Regge trajectories of Figs. 1-7 have the same universal slope of Eq.(3), which is derived with a 70 MeV quantum and the fine structure constant α = 1/137. In Table I, we show the masses and intercepts for mesons Regge trajectories of Fig. 1. For the series of mesons starting with the ρ(770), the negative vertical intercept would correspond to a non-existent pole because $J < 1$ is not allowed in an $S = 1$ state. This negative intercept is also shown in Table I. Tang and Norbury [8], likewise, studied this particular meson series and noted that this trajectory passes their defined zone test for mesons up to J = 4. However, if we were to include mesons up to J = 6, the zone test would fail. Therefore, there is a need for a better understanding of the physics involved in this particular series, which we will discuss shortly by introducing the idea of particle binding energies.

In graphing the meson Regge trajectories, we calculate error bars for several of the data by assigning an error at given point equal to 2MdM, where dM is the uncertainty in the peak mass as taken from the Particle Data Group (PDG) listing [6]. For reasons of clarity, we do not show the calculated error bars for all mesons and, later, for all baryons.

In studying Fig. 1, we note that the series beginning with ρ(770) has alternating isospin G-parity and spin parity: ρ(770)$1^+(1^{--})$, $a_2(1320)1^-(2^{++})$, $\rho_3(1690)1^+(3^{--})$, $a_4(2040)1^-(4^{++})$, $\rho_5(2350)1^+(5^{--})$, and $a_6(2450)1^-(6^{++})$. Likewise, we note a similar pattern of alternating isospin G-parity and spin parity in Fig. 2 for the series starting with ϕ(1020): namely, ϕ(1020)$0^-(1^{--})$, $f'_2(1525)0^+(2^{++})$, $\phi_3(1850)0^-(3^{--})$, and $f_J(2220)0^+(4^{++})$. In Fig. 3, we note again a similar pattern of alternating parity, beginning with ω(782): ω(782)$0^-(1^{--})$, $f_2(1270)0^+(2^{++})$, $\omega_3(1670)0^-(3^{--})$, and $f_4(2050)0^+(4^{++})$. The series ends with $f_6(2510)0^+(6^{++})$, and there is a missing meson in between. Therefore, we predict this meson to be $\omega_5(2280)0^-(5^{--})$ from the obvious pattern. In Fig. 3, we have introduced $f_J(2220)$ at J = 3, because it fits into the series for η'(958). However, this may not be correct since experiments indicate that $f_J(2220)$ better fits an assignment with J = 2 or J = 4, and we have already utilized this meson in Fig. 2.

In Fig. 4, the two meson series fit the slope of Eq.(3) with small experimental deviations which we shall discuss later. In Fig. 5, we have three series of meson Regge trajectories. For the series with $h_1(1170)$, we note again the alternating pattern of isospin G-parity and spin parity. There appears to be a missing meson at J = 3, and we predict this meson to be $h_3(2000)$. For the series with $f_1(1420)$, we extrapolate to lower mass for the intercept at J = 0 using the slope of Eq.(3), and we obtain a mass at 850 MeV. We, therefore, predict a $f_0(850)$ meson at the intercept for this particular series. There is, in fact, some evidence for this meson, which is also called the sigma meson [6]. Moreover, in Fig. 5, we do not associate the $\eta(547)$ meson with the $h_1(1170)$ meson; these mesons are located on separate Regge trajectories and do not intersect the K and $\pi$ trajectories as suggested in Fig. 10 of Tang and Norbury [8].

In Fig. 6, we show an alternate possibility for the $\pi(140)$ series compared to the series shown in Fig. 1. This series has alternating isospin G-parity; however, it does not have alternating spin parity. The mesons are along the indicated line with a slope given by Eq.(3). There are slight deviations of the experimental data from theory. No error bars are shown for this series.

Finally, we plot the meson Regge trajectories for the kaons. In Fig. 7, we have five separate series of Regge trajectories with some kaons lying below the lines with the expected slope and a few lying above the lines with the slope of Eq.(3). Overall, the patterns of kaons seem to fit the indicated lines with the slope given by Eq.(3). The series of kaons in Fig. 7 can be compared to the results of Andreev and Sergeenko [15]. The kaon series of Fig. 7 have different meson series from Andreev and Sergeenko, because these authors do not utilize the calculated slope of Eq.(3) for meson Regge trajectories.

We now discuss the apparent deviations of experimental data from theory. In Fig. 8, we plot calculations of the ratio of experimental squared masses to the theoretical as function of the angular momentum number J. The green colored line with square symbols represents the series of Fig. 3: ω(782), $f_2$(1270), $ω_3$(1670), $f_4$(2050), and $f_6$(2510). The blue colored line with diamond symbols represents the series of Fig. 1: ρ(770), $a_2$(1320), $ρ_3$(1690), $a_4$(2040), $ρ_5$(2350), and $a_6$(2450). The yellow line with triangle symbols represents the series of Fig. 7: K*(892), $K^*_2$(1430), $K^*_3$(1780), $K^*_4$(2045), and $K^*_5$(2380). The brown colored line with circle markers represents the series of Fig. 2: ϕ(1020), $f_2$(1525), $ϕ_3$(1850), and $f_J$(2220). The red colored line with cross markers represents the series of Fig. 6: π(140), $b_1$(1235), $π_2$(1670), and $ρ_3$(1990).

These ratios are normalized to the intercepts for each line at J = 0. The ratio of 1.0 is indicated with a horizontal dashed line. In the CQ model of Mac Gregor, the ratio of 1.0 would represent 0% CQ binding energy. For a 3% CQ binding energy, which is typical in the CQ model, a ratio of 0.97 would be represented by the horizontal dotted line in Fig. 8. It is apparent that a majority of the data lies outside the CQ binding energies of 0% to 3%. These sets of Regge trajectories were selected from Figs. 1 to 7 for their obviously large deviations from the theoretical lines. We note that the maximum deviations of these ratios do not exceed ± 6% from the band represented by the CQ binding energies of 0% to 3% in Fig. 8.

In Fig. 9, we again plot calculations of the ratio of experimental squared masses to the theoretical as function of the angular momentum number J. The green colored line with square symbols represents the series of Fig. 4: $a_0$(980), $ρ_1$(1450), and $ρ_3$(2250). The blue colored line with diamond symbols represents the series of Fig. 2: $f_0$(980), $f_1$(1510), and $f_2(1910)$. The yellow line with triangle symbols represents the series of Fig. 3: η'(958), $f_1(1510)$, $\eta_2(1870)$, and $f_J(2220)$. The brown colored line with circle markers represents the series of Fig. 5: $f_0(850)$, $f_1(1420)$, $\eta_2(1870)$, and $f_J(2220)$. The red colored line with cross markers represents the series of Fig. 1: π(1300), $\rho_1(1700)$, and $\pi_2(2100)$. In Fig. 9, the ratios are normalized to the intercepts for each line at J = 0. The ratio of 1.0 is indicated with a horizontal dashed line and represents a 0% CQ binding energy. For a 3% CQ binding energy, a ratio of 0.97 would be represented by the horizontal dotted line in Fig. 9. It is now apparent that more than half of the data lies inside the CQ binding energies of 0% to 3%.

Comparing the curves of Figs. 8 and 9, we may ask what are the dynamics of the series in Fig. 8, which produce data outside the expected CQ binding energy range. A possible explanation again comes from the CQ model of Mac Gregor [9]. Mac Gregor has noted that a meson like ρ(770) has a broad width and is a rotational excitation. Thus, it must have a range of angular momentum values (cf. pages 1302-1303 of Ref. 9). The ρ meson has a threshold mass given by [7]:

$$E_{\text{threshold}} = E_{\text{peak}} - \Gamma_{\text{FWHM}}. \quad (4)$$

Eq.(4) indicates that the ρ meson has a threshold mass of about 617 MeV. Knowing this information, we can make a correction to the blue line of Fig. 8. This adjustment is shown as the black curve with open circles in Fig. 10 and is tabulated in Table II. The blue curve with diamond symbols represents the original, uncorrected data. The majority of data, represented by the black curve, now lies inside the CQ binding energy range as indicated by the two horizontal lines. Although this procedure works well for the ρ meson and its series, we have not investigated the remaining curves of Fig. 8 for possible adjustments. The author recommends further investigation by the readers.

As a final note about meson Regge trajectories, it is known that the slopes for the D mesons, the charmonium states, and the bottomonium states are divergent in comparison to those in Figs. 1-7, as noted by Tang and Norbury [8]. It is expected that the D mesons would have heavier quark masses. A larger quark mass would produce a larger slope for a particular meson series. This idea is consistent with Fig. 10 of Tang and Norbury [8]. However, caution must be taken in identifying the correct order of a particular meson series. Two data points do not necessary define the correct Regge trajectory as shown in Fig. 10 of Tang and Norbury [8] nor in our Figs. 1-7. Therefore, the application of Regge theory to heavier mesons is suspect when there is a scarcity of data.

## III. BARYON REGGE TRAJECTORIES

In a comprehensive study of Regge trajectories, Inopin [3] has reviewed several relativistic and semi-relativistic models. Inopin noted that some authors claimed that baryon Regge slopes are noticeably small compare to the meson Regge trajectories and that a quark-diquark structure for baryons cures this defect. In fact, Berdnikov and Pron'ko [16] proposed such a cure with a relativistic quark-diquark model and stated, "the slopes of the baryonic trajectories practically coincide with those of the mesonic trajectories, which is in favor of the quark-diquark structure." With the large error bars shown in Figs. 6 and 7 of Berdnikov's and Pron'ko's work [16], there is some suspect to the claim that the baryon Regge slopes are coincident with the meson Regge slopes. There is no *a priori* reason why the slopes should be the same for both meson and baryon Regge trajectories. Mesons are two quark systems, and baryons are 3-body systems. If the slopes are proportional to the quark masses and to the coupling constant, then there are more than likely differences in the slopes for mesons and baryons.

We note that the slope was derived for meson Regge trajectories, Eq.(2), from Barut's solution to a relativistic Balmer mass formula. In a like manner, the slope for the baryon Regge trajectories should be derived as:

$$\text{Slope (baryons)} = [(m_1 m_2 m_3)/(m_1 + m_2 + m_3)]\ \alpha_S, \qquad (5)$$

where $m_1$, $m_2$, and $m_3$ are the individual quark masses, and $\alpha_S$ is the strength of the coupling constant.

We again turn to the CQ model of Mac Gregor for the selection of the individual quark masses. If the nucleon is the start of a series for baryon Regge trajectories, then the individual quark masses must be approximately $m_1 = m_2 = m_3 = u = 315$ MeV for the u-quark. The masses of the u and d quarks are comparable to each other. However, in the CQ model the particle masses are normally less 3% binding energy, so that we choose the quark masses to be (0.97)(315) = 305.6 MeV. The question remains what to select for the coupling constant $\alpha_S$. The work of Sawada has been overlooked for years when it comes to studying p-p scattering at low energies [17-20]. The strong coupling constant is [20]:

$$\alpha_S = 137/4 = 34.25 \qquad (6)$$

We note the appearance of the fine structure constant or the number 137 in Eq.(6).

Substituting the result of Eq.(6) and the reduced u-quark mass 305.6 MeV into Eq.(5), we have the following for the slope of the baryon Regge trajectories:

$$\text{Slope (baryons)} = (0.3056)^2(1/3)(137/4) = 1.0662\ \text{GeV}^2\ . \qquad (7)$$

Thus, we have derived a slope for the baryon Regge trajectories, which is less than the slope of 1.3426 $GeV^2$ for the meson Regge trajectories.

We now reproduce the spectra of baryons for masses below 3000 MeV which are dependent upon the slope of Eq.(7). These baryon Regge trajectories are shown in Figures 11 to 16. The baryon Regge trajectories of Figs. 11-16 have the same universal slope of Eq.(7), which is derived with a reduced quark mass, as in the CQ model, and with the fine structure constant or the number 137. In Fig. 11, the squared masses of the nucleons are on lines with the slope of Eq.(7), and the series starting with N(939) is shown as a solid line. For the N(1440) series, as shown by the dotted lines, there is a deviation of the squared mass for the N(1440) data point from the expected slope of Eq.(7). In Fig. 12, the delta baryons fit the lines with the expected slope. From the intercept of the dotted line at J = ½, we have predicted the existence of Δ(1079). This baryon was predicted in an earlier paper [21]. For the data in Fig. 13, there is a clear fit to the slope of Eq.(7), starting with Λ(1116) in the series noted by the dashed line. However, we note the deviation of the series marked by the initial states at Λ(1405) and Λ(1600). There is a scarcity of data for the Λ(1405) series.

In Fig. 14, we compare the slopes of the lines for both Λ(1116) and N(939). It is clear that the slopes of these two lines are the same as that derive in Eq.(7). In Fig. 15, we have plotted the data for the sigma baryons. The series, starting with Σ(1192), fits the slope of Eq.(7), as indicated by the dashed line. However, we clearly see deviations from the expected slope of other data points in Fig. 15, but the overall trend is consistent with the slope of Eq.(7). The series beginning with Σ(1750) appears to be divergent, according to the criteria proposed by Tang and Norbury [8]. In Fig. 16, we show the data for Ξ and Ω baryons together. The series for the Ξ baryons is also consistent with a slope found in Eq.(7). For the Ω baryons, we have extrapolated the line to J = ½ and J = 3/2, predicting the existence of Ω(1570) at J = ½ and Ω(1868) at J = 3/2. Finally, as we noted in the previous section on the meson Regge trajectories, we can expect the slope of baryon Regge trajectories to diverge from the derived slope in Eq.(7) if the quark masses increase due a change in quark flavor (e.g., change to a charm or bottom quark).

## IV. CONCLUSION

In this paper, we presented a model by Barut [4, 5] and derived equations for the slopes of both meson and baryon Regge trajectories. We established *different* slopes for meson and baryon Regge trajectories from the data published by the Particle Data Group [6], and showed how the meson Regge trajectories are dependent upon a 70 MeV quantum proposed by Mac Gregor [7]. We compared our results with the recent review by Tang and Norbury [8] on the properties of Regge trajectories. It was found that meson Regge trajectories have a universal slope of 1.3426 $GeV^2$, which is proportional to the fine structure constant $\alpha$ or to the number 137. Likewise, we derived a slope of 1.0662 $GeV^2$ for the baryon Regge trajectories, which is less than the slope for the mesons. The slope of the baryon Regge trajectories is also proportional to the fine structure constant. Although the theoretical formulas showed good agreement with the experimental data, there are some obvious deviations of the data from theory. We attributed some of these deviations as due to constituent-quark (CQ) binding energies, as modeled by Mac Gregor. Other contributions from spin-dependent forces will no doubt add to our understanding of these experimental deviations from theory.

TABLE I. Masses and intercepts for mesons Regge trajectories in Fig. 1 are derived with mass formula $M^2 = m^2 + (1.3426)J$ and with the universal slope of Eq.(3). For a given series, the squared masses are in units of $GeV^2$ and particles are identified where applicable.

| Meson parent | J = 0 | 1 | 2 | 3 | 4 | 5 | 6 |
|---|---|---|---|---|---|---|---|
| | | | | $m^2$ ($GeV^2$) | | | |
| $\pi(140)$ | 0.0196 | 1.3622 | 2.7048 | 4.4047 | 5.390 | | |
| | $\pi(140)$ | $a_1(1260)$ | $a_2(1700)$ | $\rho_3(1990)$ | | | |
| $\pi(1300)$ | 1.5 | 2.8426 | 4.1852 | | | | |
| | $\pi(1300)$ | $\rho(1700)$ | $\pi_2(2100)$ | | | | |
| $\rho(770)$ | -0.7497 | 0.5929 | 1.9355 | 3.278 | 4.620 | 5.963 | 7.3059 |
| | | $\rho(770)$ | $a_2(1320)$ | $\rho_3(1690)$ | $a_4(2040)$ | $\rho_5(2350)$ | $a_6(2450)$ |

TABLE II. Masses and intercepts for mesons Regge trajectories in Fig. 10 are derived with mass formula $M^2 = m^2 + (1.3426)J$ and with the universal slope of Eq.(3). The squared masses are in units of $GeV^2$ and particles are identified where applicable.

| Meson parent | J = 0 | 1 | 2 | 3 | 4 | 5 | 6 |
|---|---|---|---|---|---|---|---|
| | | | | $m^2$ ($GeV^2$) | | | |
| $\rho(770)$ | -0.7497 | 0.5929 | 1.9355 | 3.278 | 4.620 | 5.963 | 7.3059 |
| | | $\rho(770)$ | $a_2(1320)$ | $\rho_3(1690)$ | $a_4(2040)$ | $\rho_5(2350)$ | $a_6(2450)$ |
| $\rho(617)$ | -0.9457 | 0.3969 | 1.7398 | 3.084 | 4.4226 | 5.770 | 7.1076 |
| | | $\rho(617)$ | $a_2(1320)$ | $\rho_3(1690)$ | $a_4(2040)$ | $\rho_5(2350)$ | $a_6(2450)$ |

## FIGURE CAPTIONS

**Fig. 1.** Regge trajectories are shown for *three* separate series of mesons with a universal slope given by Eq.(3) in the text. The data represented by the dotted line fits a series which has alternating isospin G-parity and spin parity, starting with the ρ(770) meson at J = 1.

**Fig. 2.** Regge trajectories are shown for *five* separate series of mesons with a universal slope given by Eq.(3) in the text. The data represented by the dotted line fits a series which has alternating isospin G-parity and spin parity, starting with the ϕ(1020) meson at J = 1.

**Fig. 3.** Regge trajectories are shown for *two* separate series of mesons with a universal slope given by Eq.(3) in the text. The data represented by the dotted line fits a series which has alternating isospin G-parity and spin parity, starting with the ω(782) meson at J = 1.

**Fig. 4.** Regge trajectories are shown for *two* separate series of mesons with a universal slope given by Eq.(3) in the text.

**Fig. 5.** Regge trajectories are shown for *three* separate series of mesons with a universal slope given by Eq.(3) in the text. The data represented by the dotted lines fits a series which has alternating isospin G-parity and spin parity, starting with the $h_1$(1170) meson at J = 1.

**Fig. 6.** An alternate series of mesons is shown on a Regge trajectory with a slope given by Eq.(3) in the text.

**Fig. 7.** Regge trajectories are shown for *five* separate series of K mesons with a universal slope given by Eq.(3) in the text.

**Fig. 8.** Calculations are shown as a function of J for the deviations of the experimental data from theory. The horizontal dashed line represents 0% CQ binding energy. The horizontal dotted line represents a 3% CQ binding energy.

**Fig. 9.** Calculations are shown as a function of J for the deviations of the experimental data from theory. The horizontal solid line represents 0% CQ binding energy. The horizontal dotted line represents a 3% CQ binding energy.

**Fig. 10.** Adjustment is made to the meson Regge trajectory, starting with the ρ(770) meson. The blue curve represents the original curve found in Fig. 8. The black curve represents the procedure for reducing the mass of the ρ meson and refit to the series with the universal slope given by Eq.(3).

**Fig. 11.** Regge trajectories are shown for *four* separate series of nucleon baryons with a universal slope given by Eq.(7) in the text.

**Fig. 12.** Regge trajectories are shown for *five* separate series of delta baryons with a universal slope given by Eq.(7) in the text. The Δ(1079) is predicted to exist.

**Fig. 13.** Regge trajectories are shown for *four* separate series of lambda baryons with a universal slope given by Eq.(7) in the text. Note the deviations from the expected slopes for a few of these baryons.

**Fig. 14.** A comparison is shown of the nucleon Regge trajectory with the lambda Regge trajectory; each trajectory has a universal slope given by Eq.(7).

**Fig. 15.** Regge trajectories are shown for *five* separate series of sigma baryons with a universal slope given by Eq.(7) in the text. Note the deviations from the expected slopes for several of these baryons.

**Fig. 16.** Regge trajectories are shown for *two* separate series of baryons with a universal slope given by Eq.(7) in the text. The Ω(1570) at $J = ½$ and Ω(1868) at $J = 3/2$ are predicted to exist.

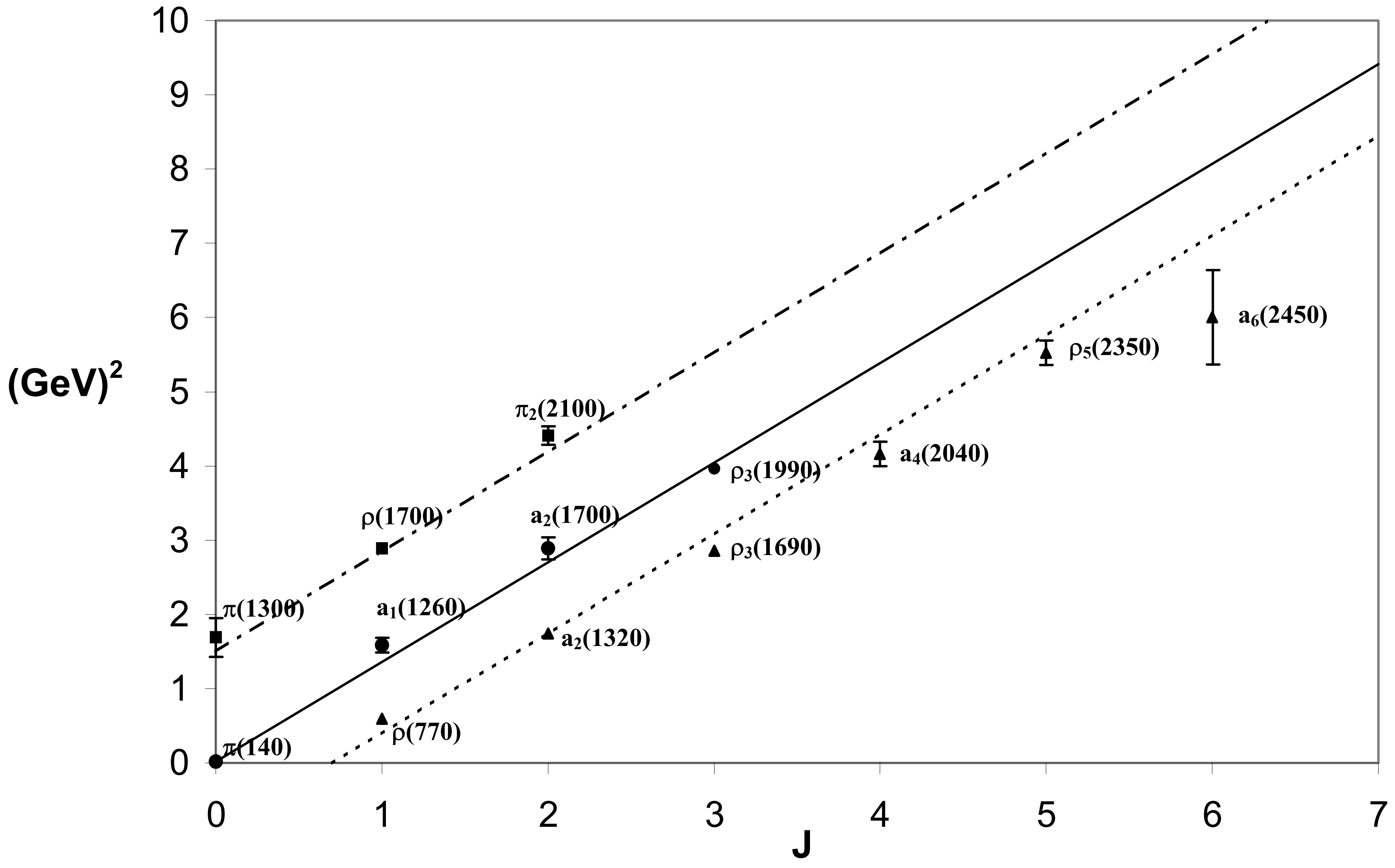

**Fig. 1.**

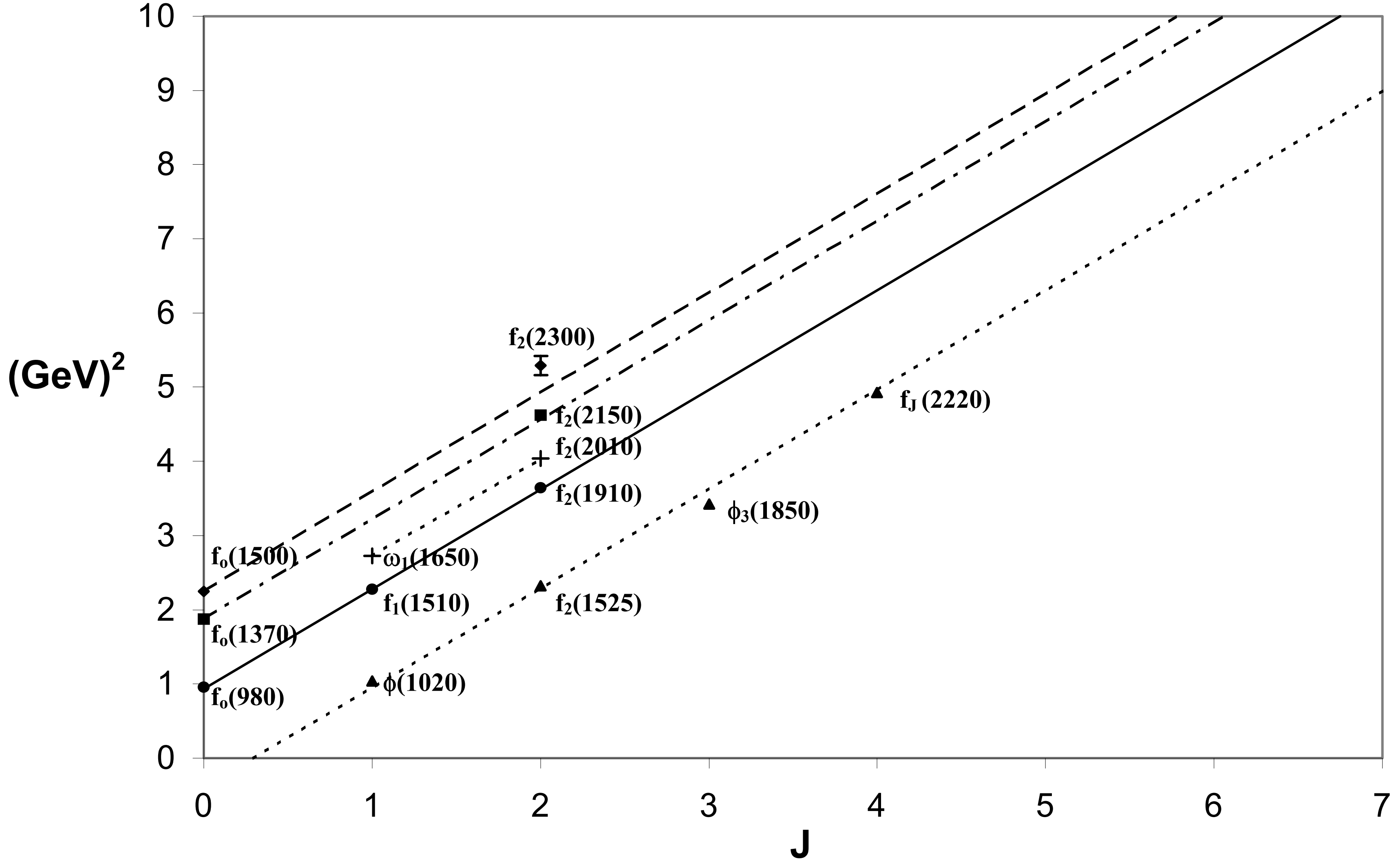

Fig. 2.

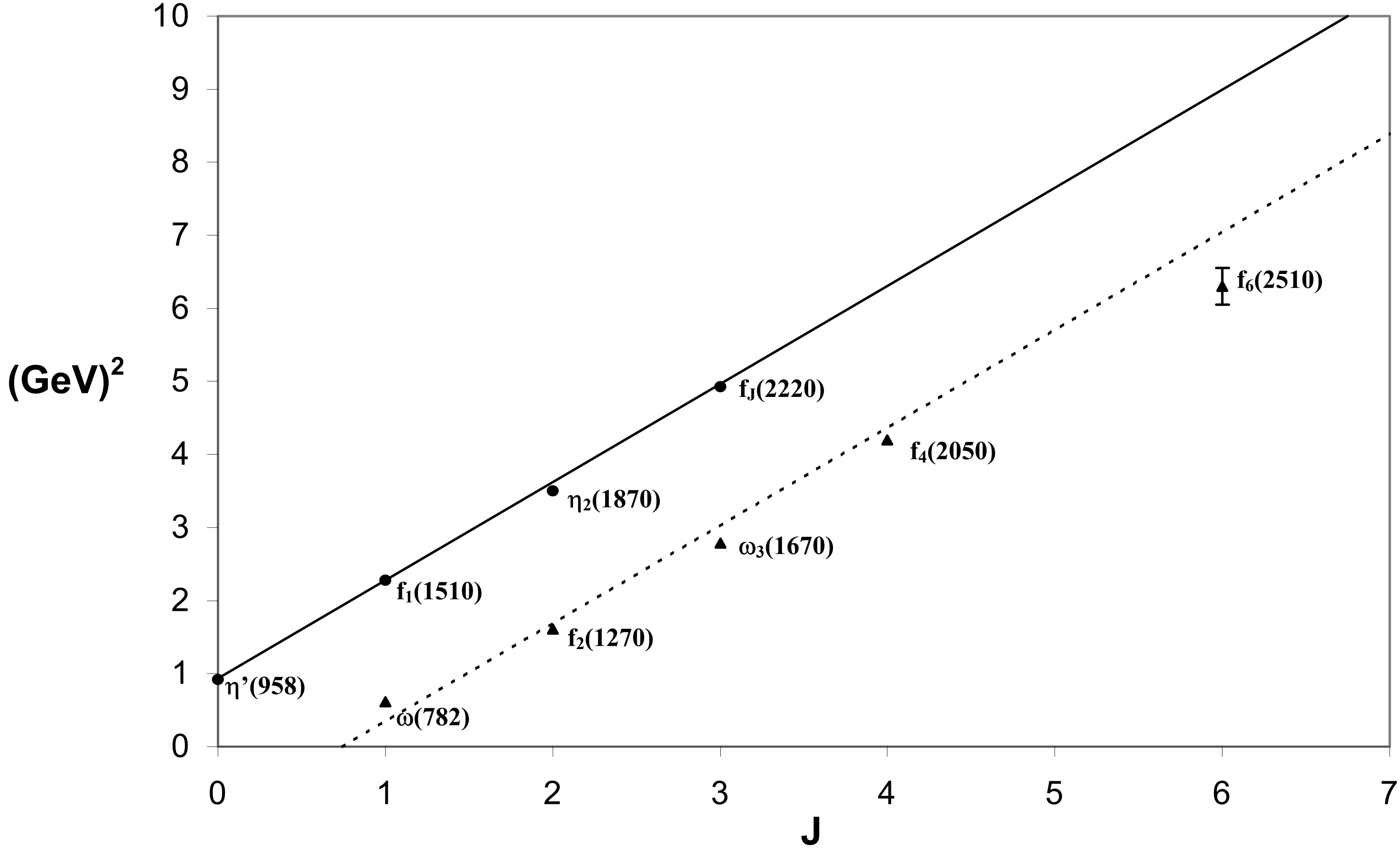

**Fig. 3.**

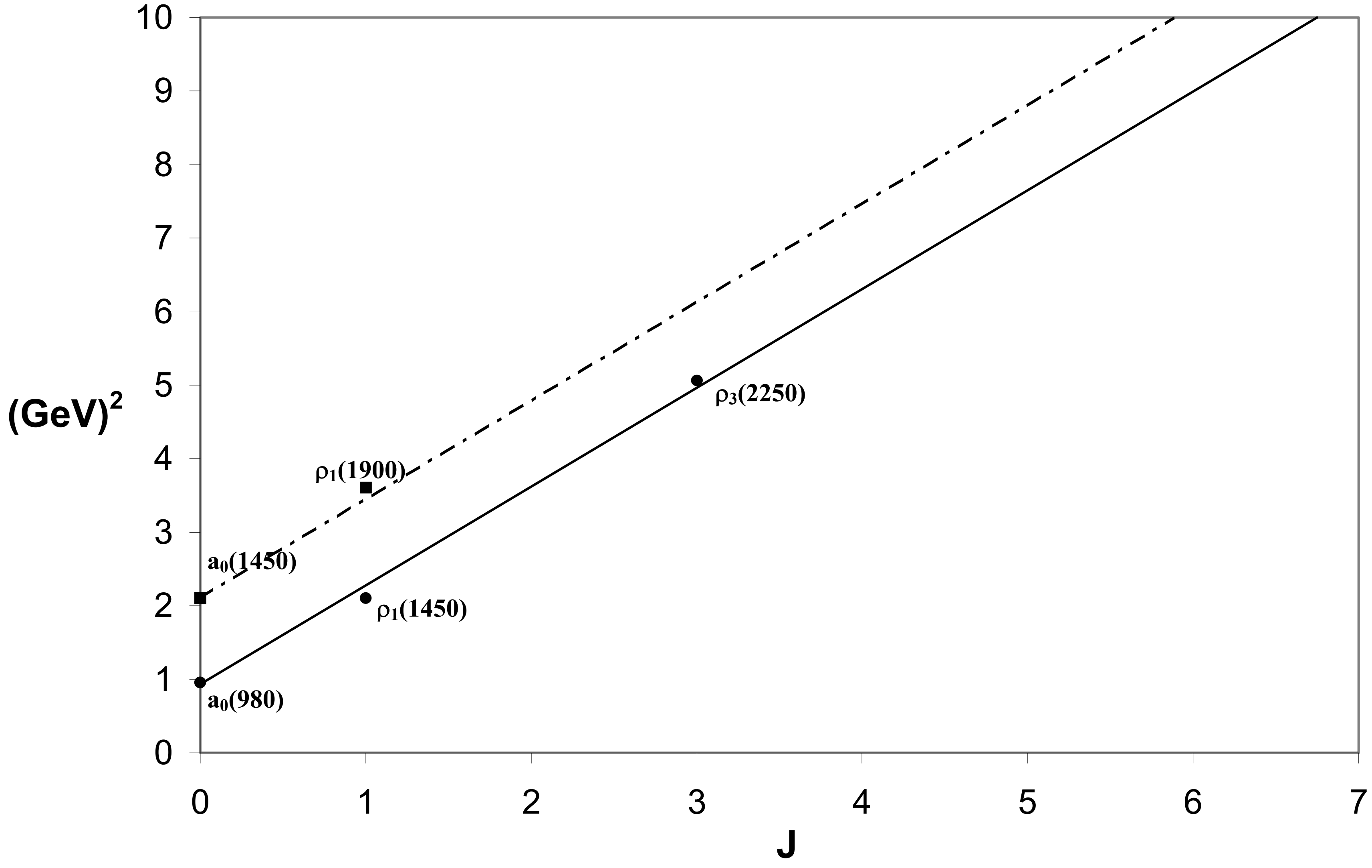

**Fig. 4.**

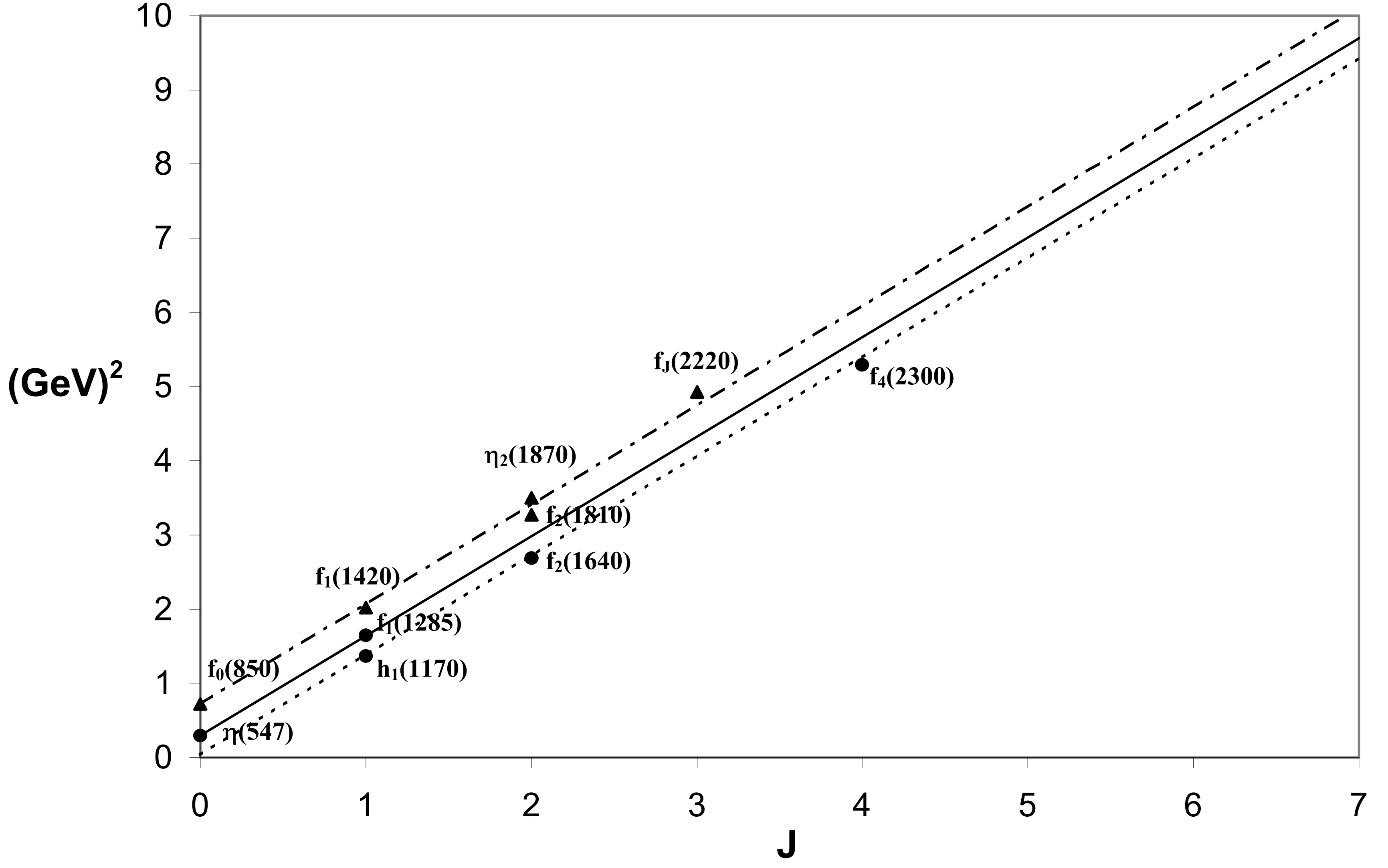

Fig. 5.

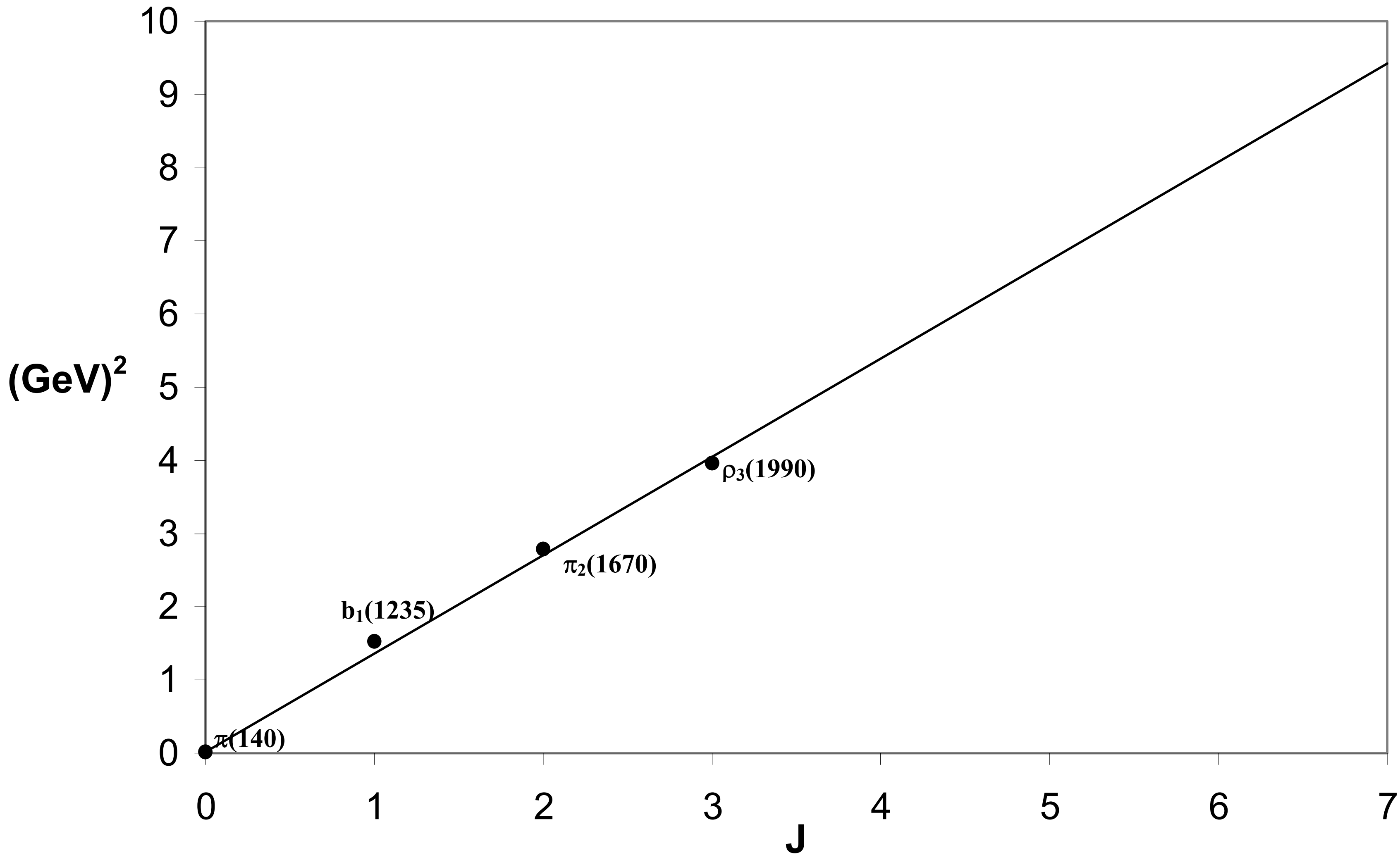

Fig. 6.

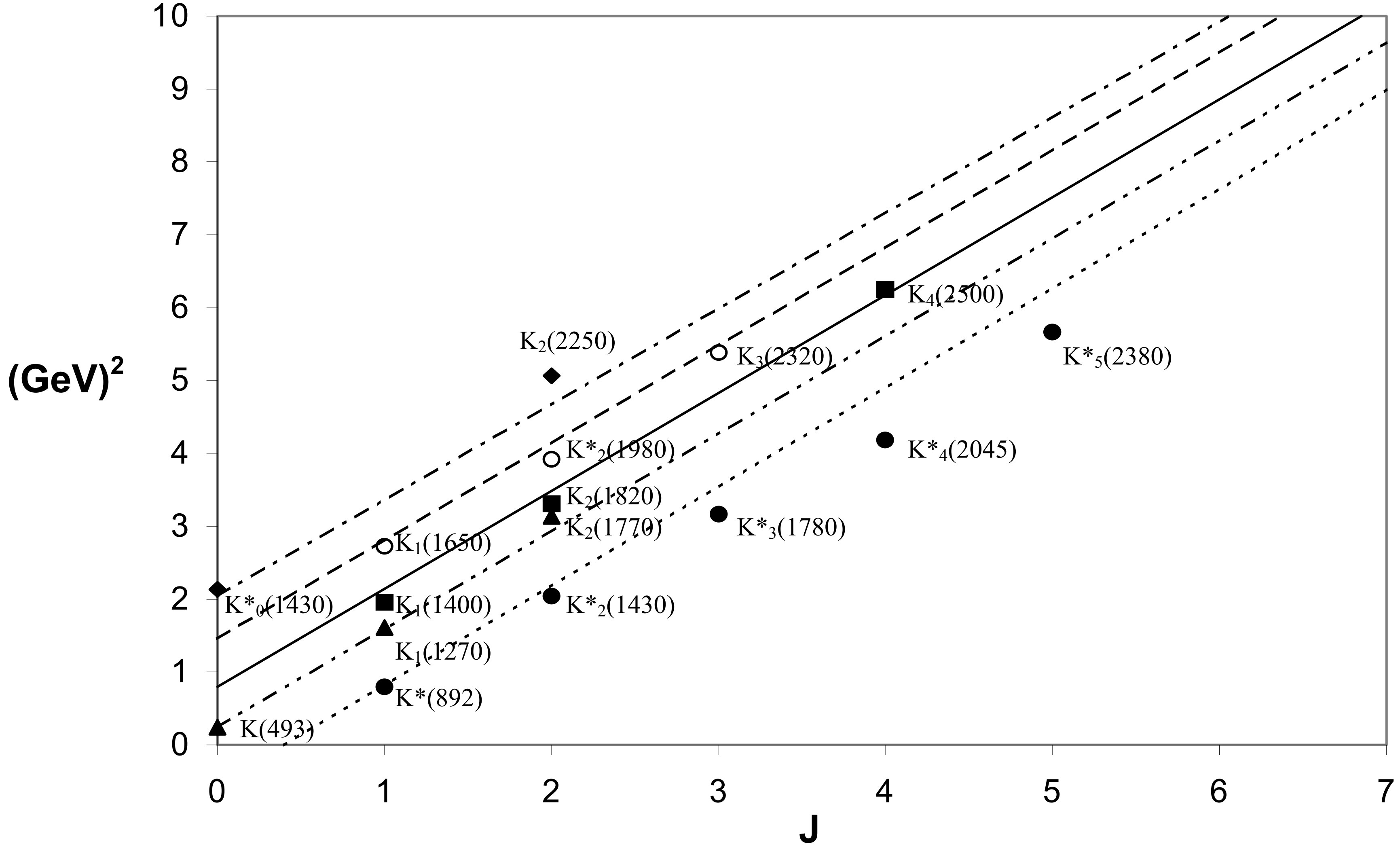

Fig. 7.

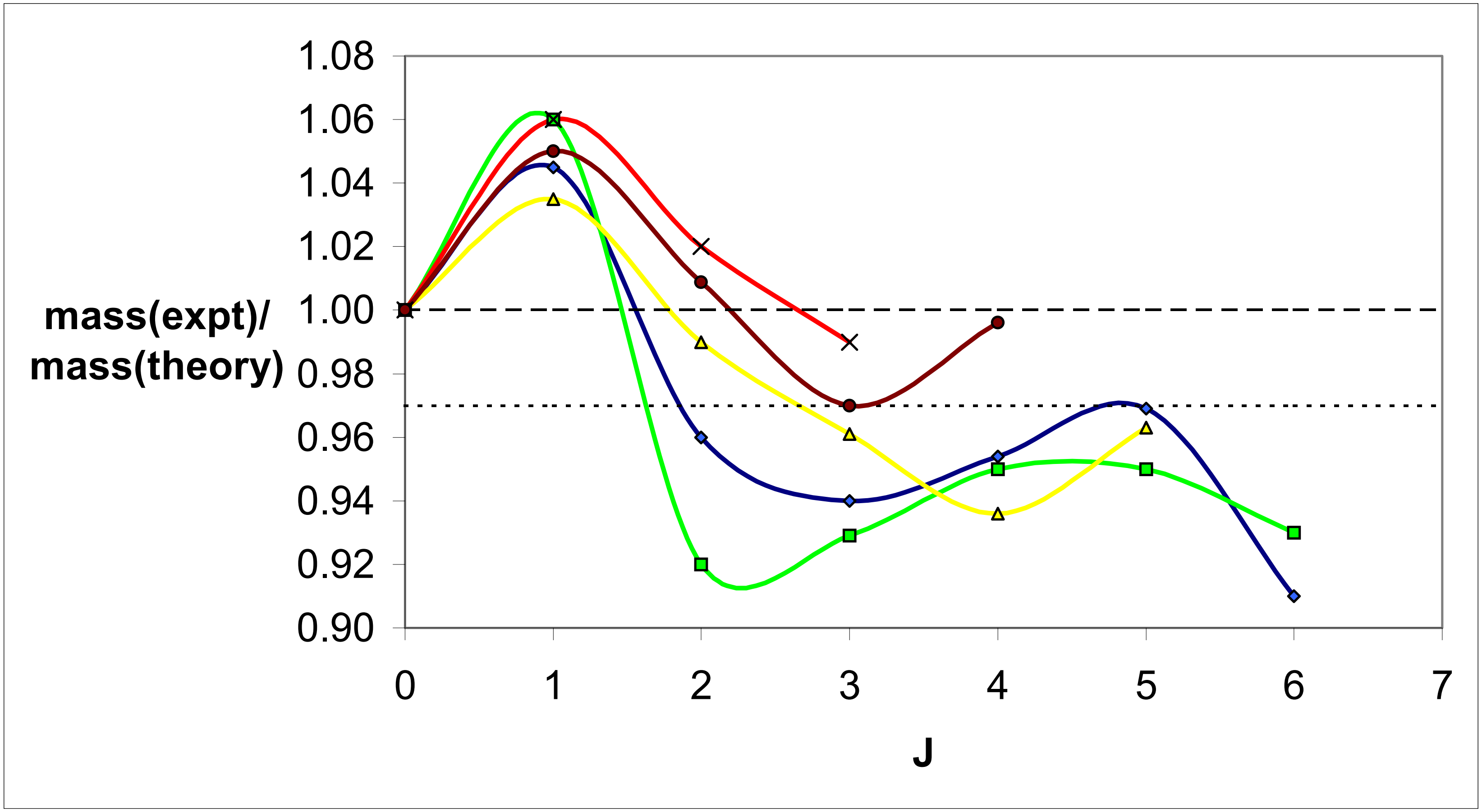

Fig. 8.

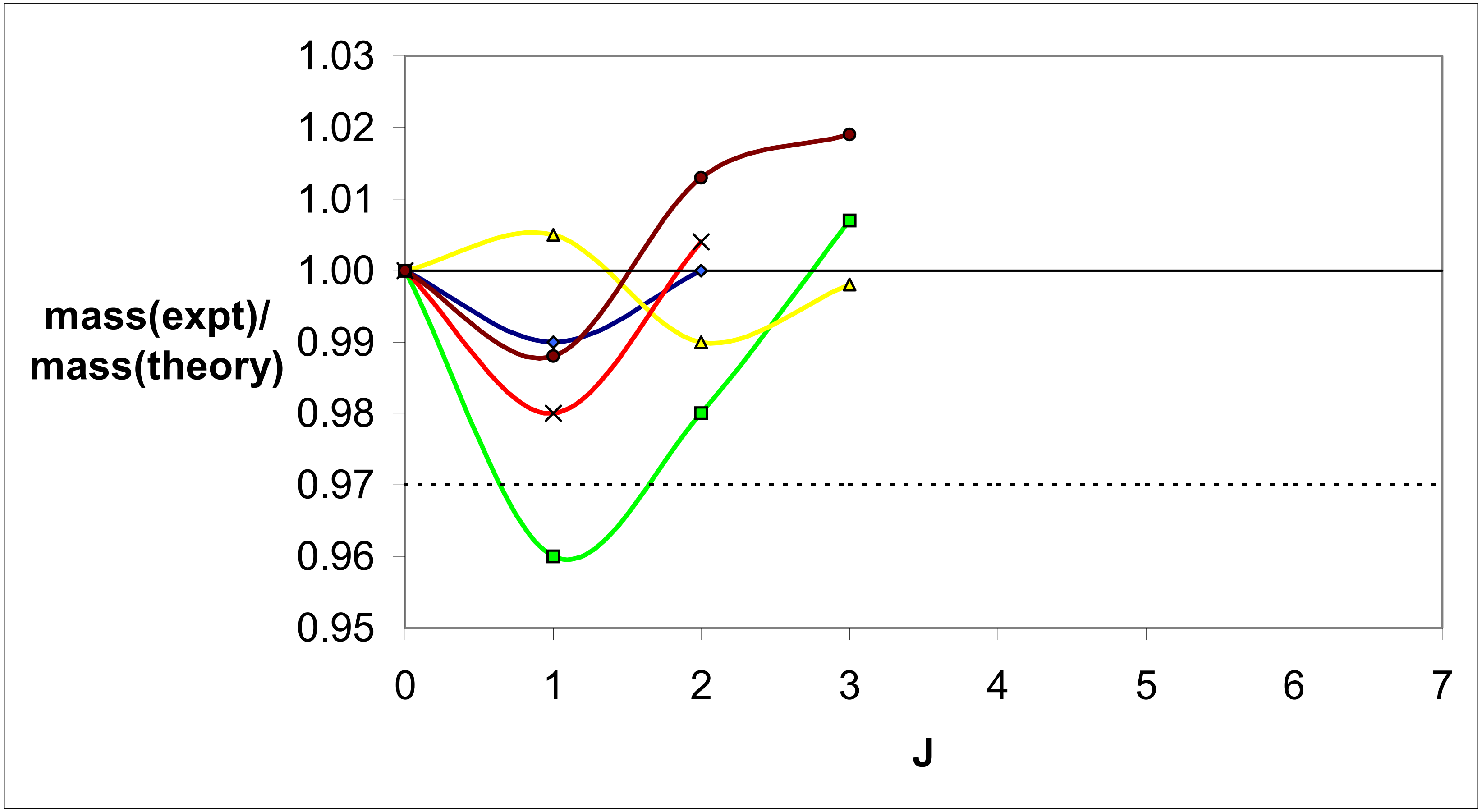

**Fig. 9.**

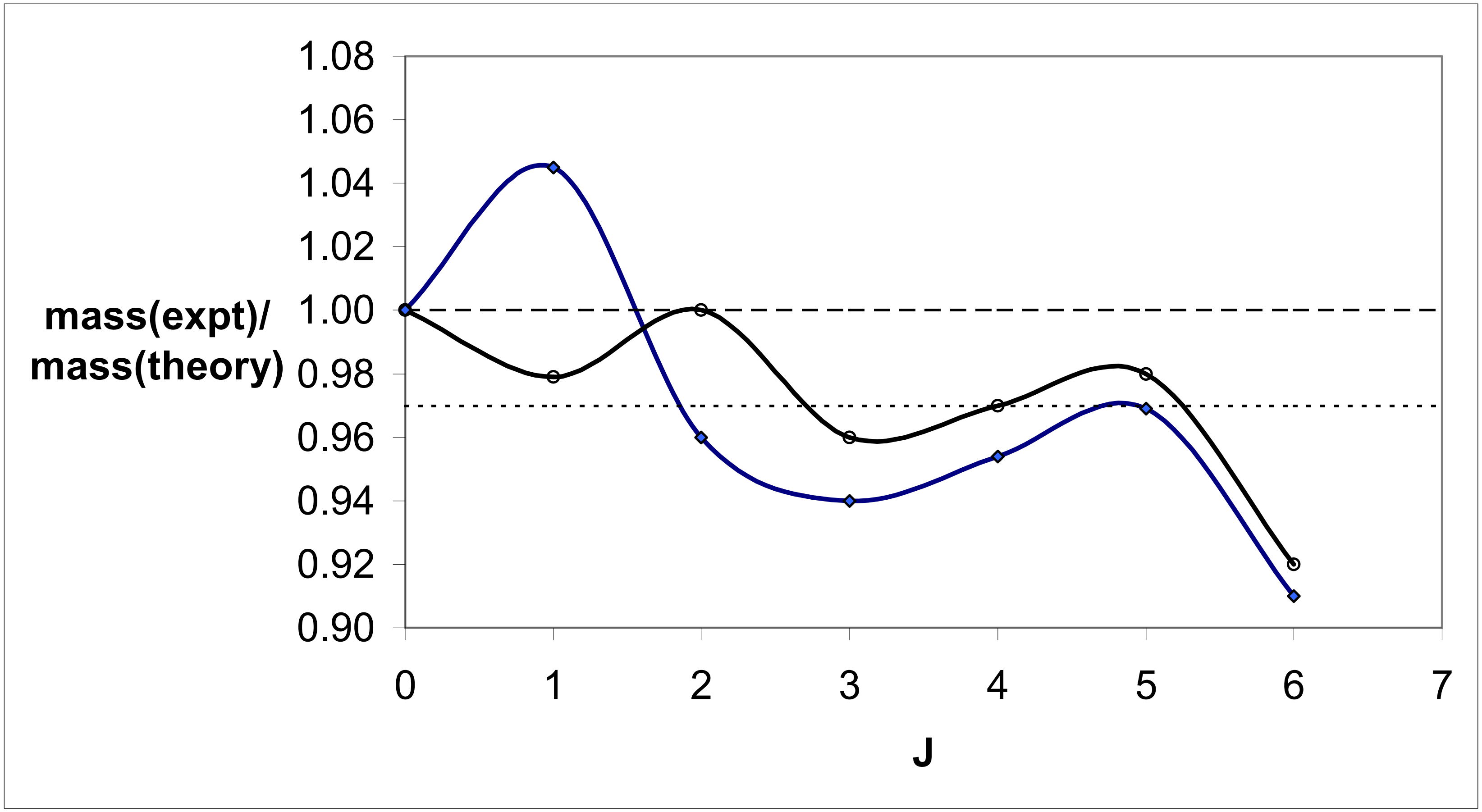

**Fig. 10.**

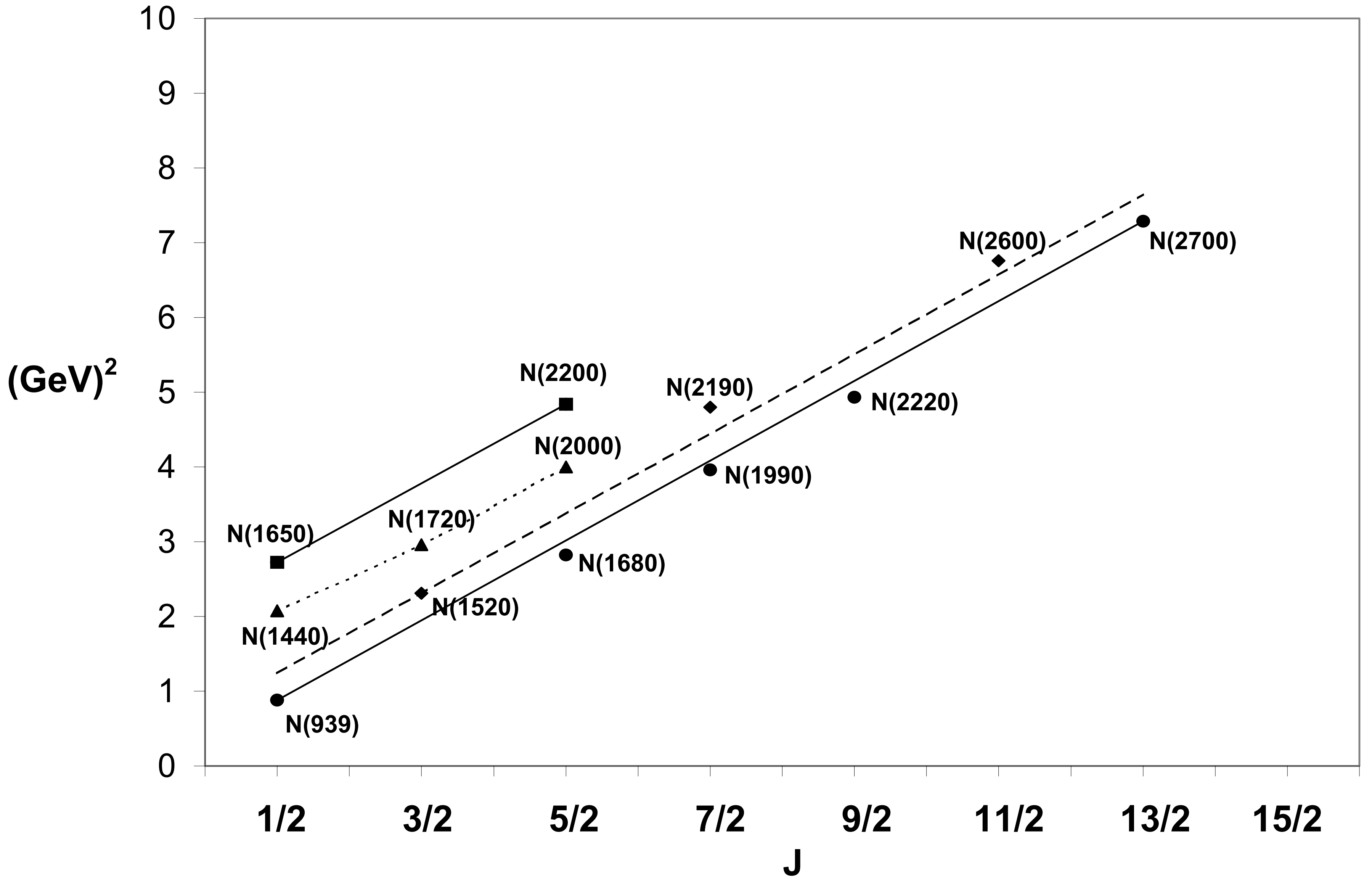

Fig. 11.

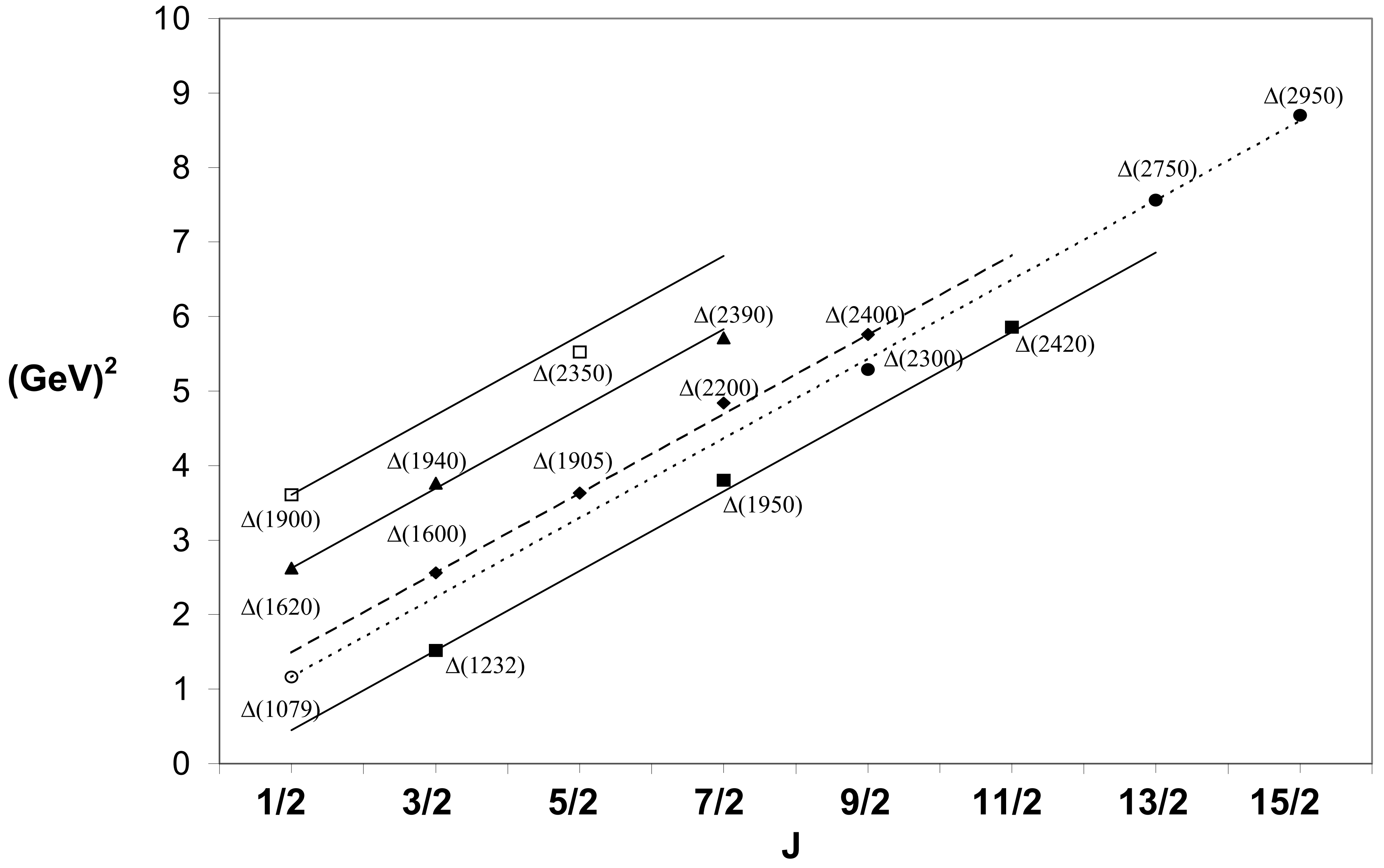

**Fig. 12.**

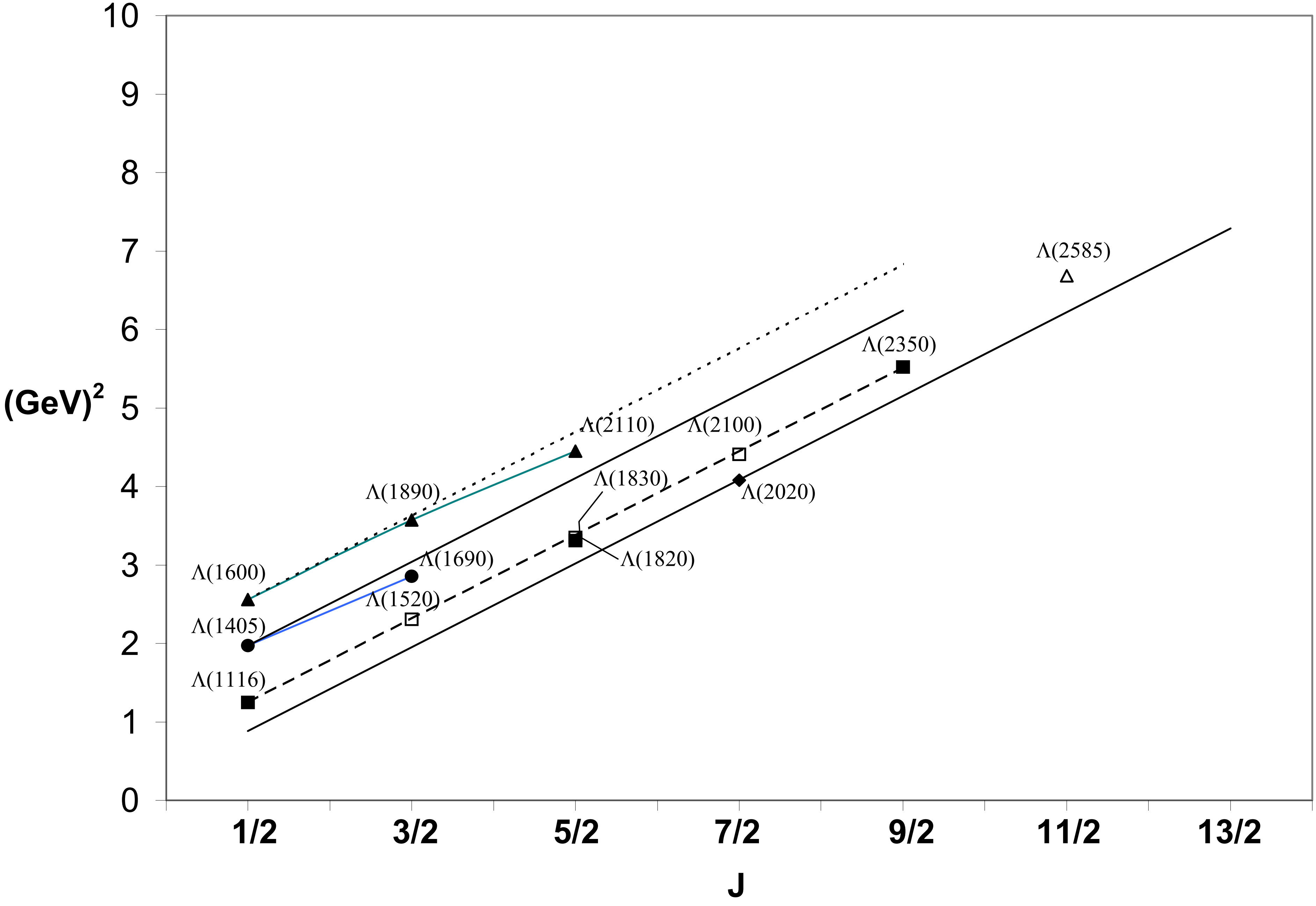

Fig. 13.

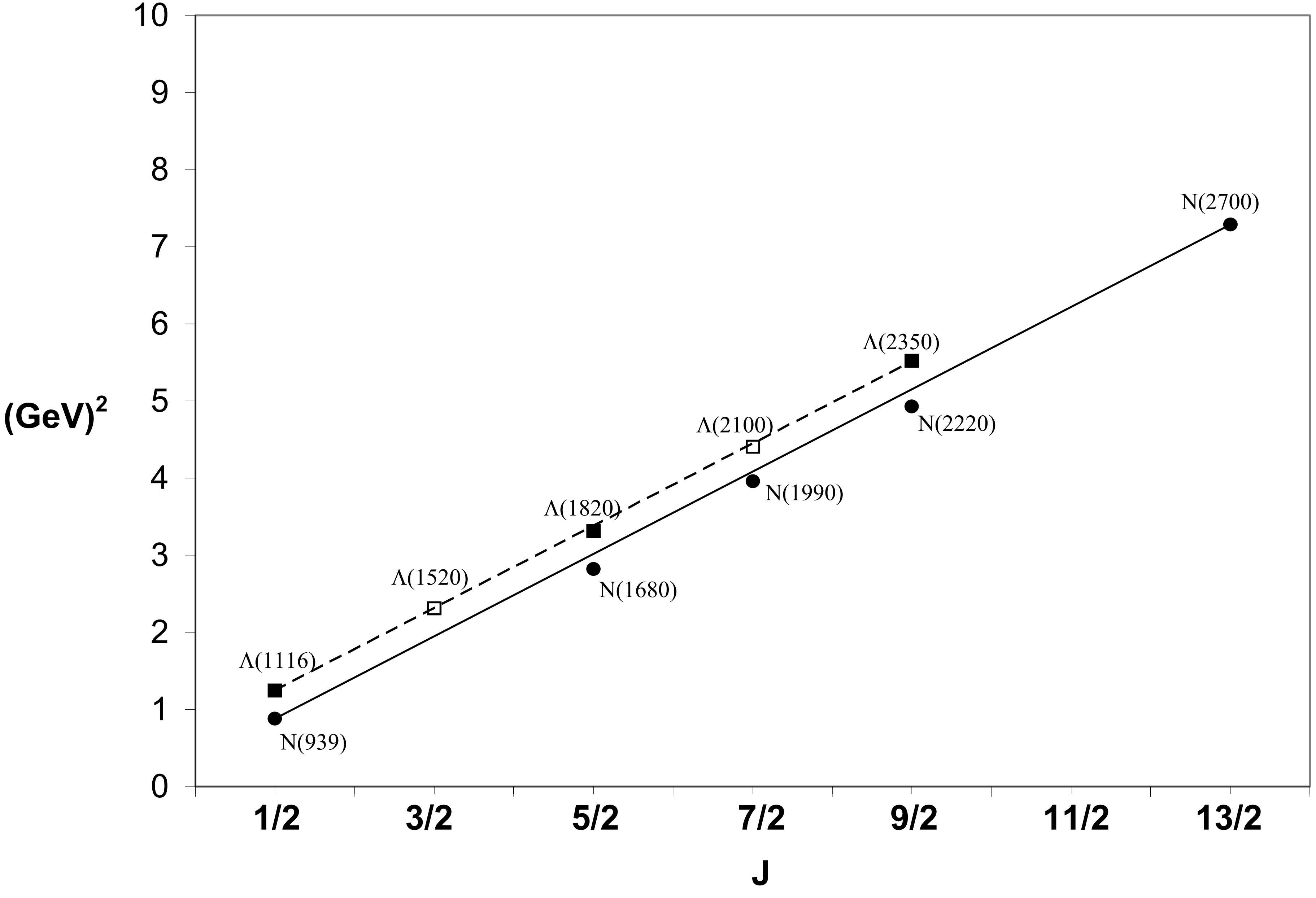

Fig. 14.

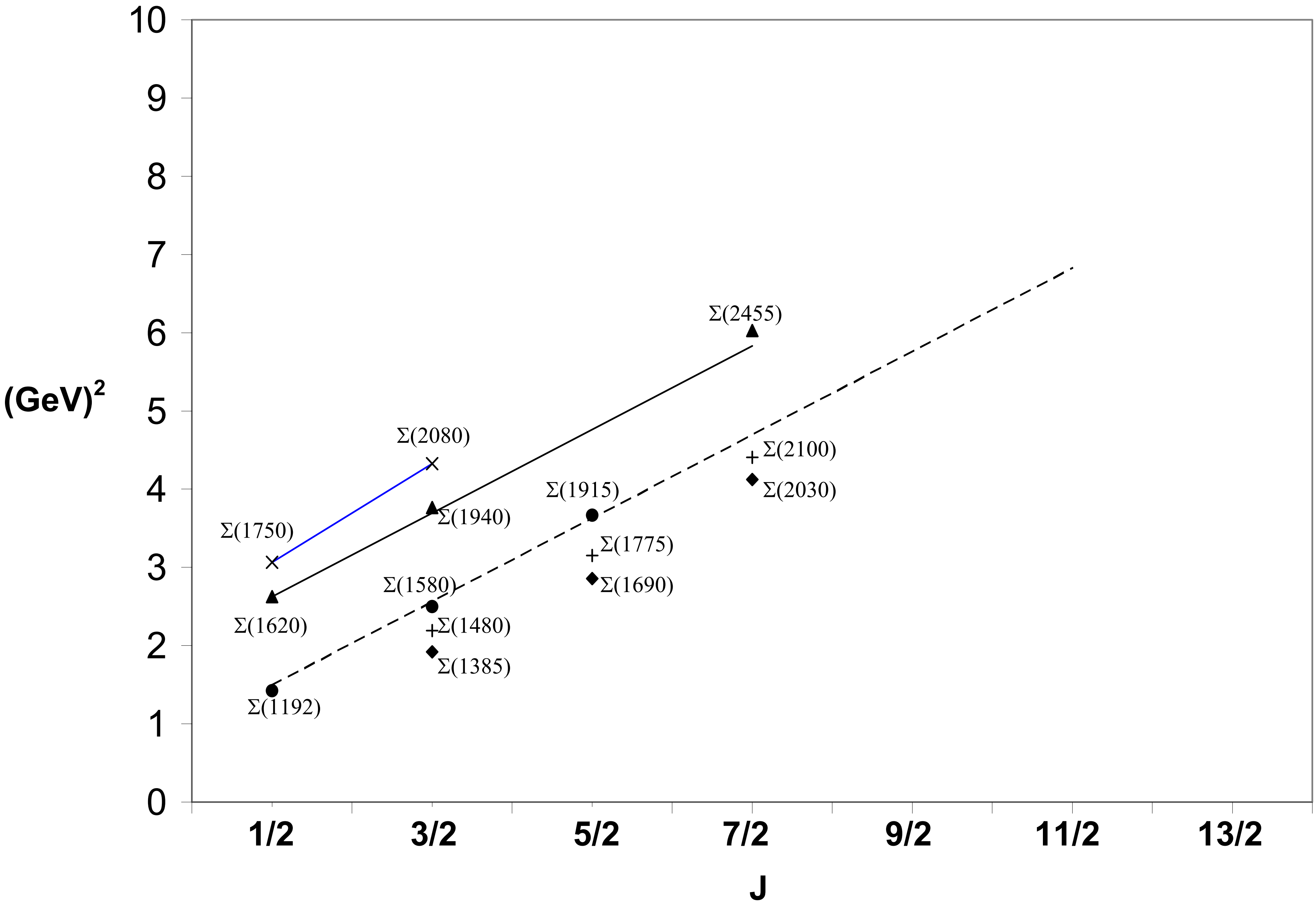

**Fig. 15.**

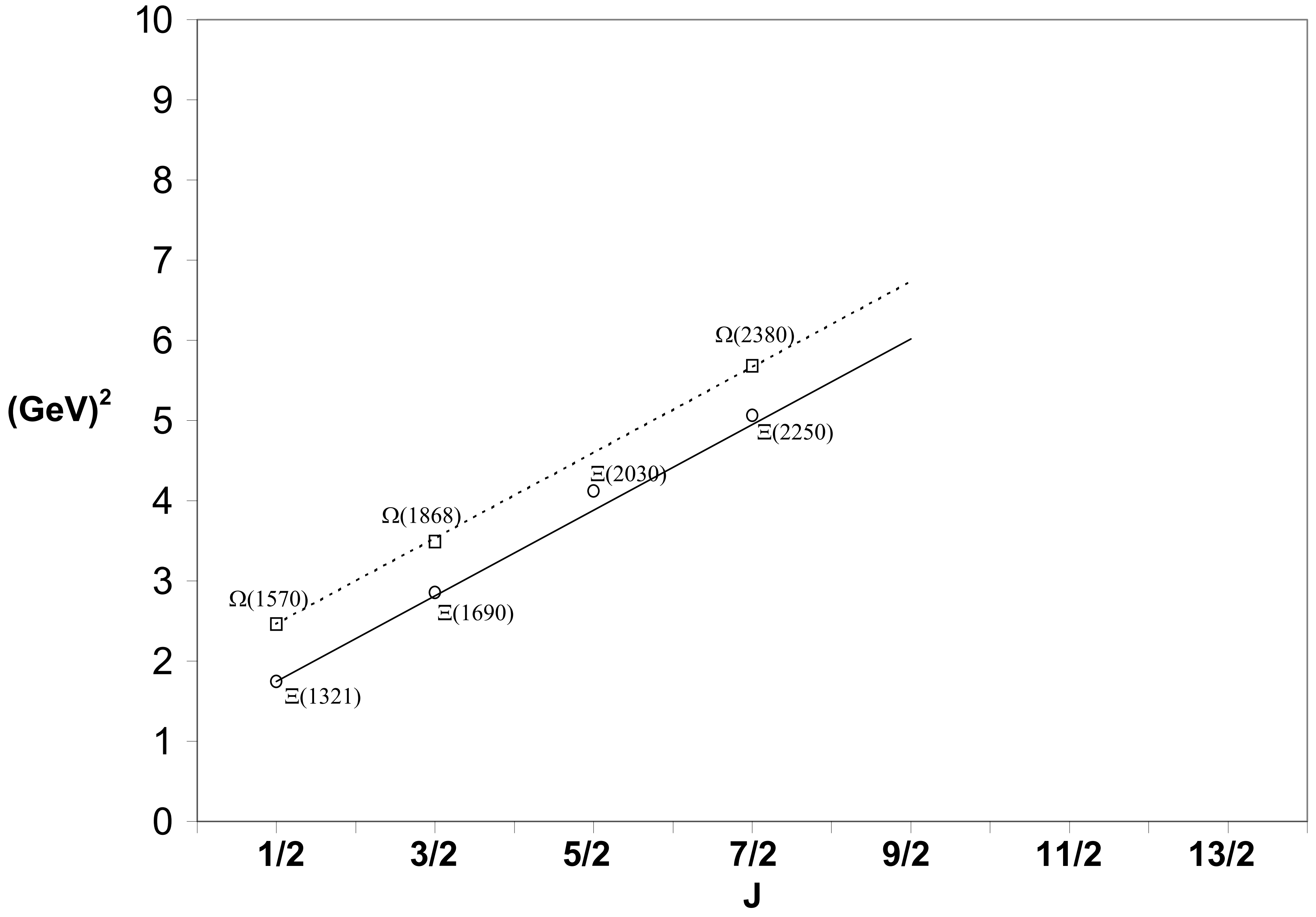

Fig. 16.